\documentclass[prl,showpacs,twocolumn]{revtex4}
\usepackage{psfig}
\begin{document}
\title{
Magnetism in Atomic-Sized Palladium Contacts and Nanowires 
}

\author{A. Delin}
\affiliation{Abdus Salam International Center for Theoretical Physics (ICTP), Strada Costiera 11, 34100 Trieste, Italy}
\author{E. Tosatti}
\affiliation{Abdus Salam International Center for Theoretical Physics (ICTP), Strada Costiera 11, 34100 Trieste, Italy\\
             International School for Advanced Studies (SISSA), via Beirut 2--4, 34014 Trieste, Italy\\
	                  INFM DEMOCRITOS National Simulation Center, via Beirut 2--4, 34014 Trieste, Italy
			               }
\author{R. A. Weht}
\affiliation{Comisi\'on Nacional de Energ\'\i a At\'omica (CNEA), 
Avda. General Paz y Constituyentes, 1650 San Mart\'{\i}n, Argentina}

\date{\today}

\begin{abstract}
We have investigated Pd nanowires theoretically, and found 
that, unlike either metallic or free atomic Pd, they exhibit 
Hund's rule magnetism. In long, monoatomic wires, we find a  spin moment of 0.7~$\mu_B$ per atom, 
whereas for short, monoatomic wires between bulk leads, the predicted moment is about 0.3~$\mu_B$ per wire atom.
In contrast, a coaxial (6,1) wire was found to be nonmagnetic.
The origin of the wire magnetism is analyzed. 
\end{abstract}


\maketitle

Magnetism at the nanoscale is an exciting emerging research field, of
both basic and applied relevance.
In future technology, when the 
electronic components will have become so tiny 
that quantum size effects will determine their functionality, 
an understanding of nanomagnetic phenomena will be crucial;
yet, relatively little is currently understood about how 
magnetism arises and how it affects the properties of metals at 
the nanoscale.

\begin{figure}[h]
\psfig{figure=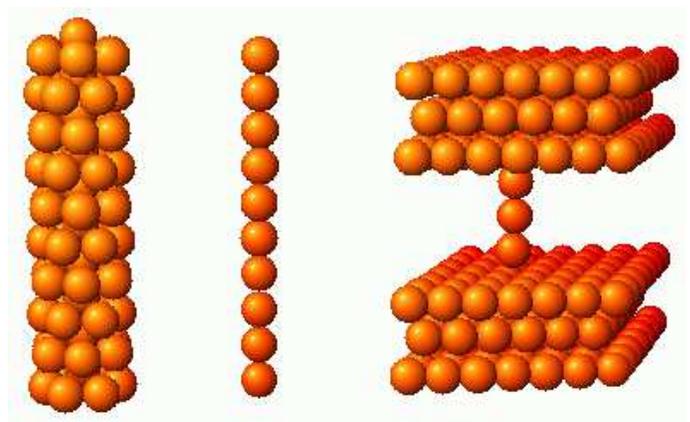,width=9.0cm}
 \caption
  {
  Schematical drawing of the wire geometries addressed in this paper. From left to right: 
  (6,1) coaxial wire; monowire; short monowire between 
  bulk leads.
 \label{fig:wire_geometries}
  }
 \end{figure}

Systems of special interest are nanowires and atomic-sized 
nanocontacts. The low dimensionality of 
such systems causes specific physical phenomena to appear, 
for example quantized ballistic conductance~\cite{wees1988} and helical wire geometries~\cite{gulseren1998}.
These phenomena interplay with the possible
presence of magnetism in the nanosystem, especially of a 
genuine Hund's rule magnetic order parameter. 
Here, we report theoretical studies of emerging magnetism in Pd nanowires in various geometries, see Fig.~\ref{fig:wire_geometries}. 
Monowires of Pd, i.e., wires consisting of a single line of atoms, have recently 
been observed by Rodrigues {\it et al.}\cite{rodrigues}.
We find magnetic moments as high as 0.7 $\mu_{\rm B}$ per atom
for infinitely long, straight, monatomic wires. 
Even short, three-atom Pd wire chains between bulk-like leads, 
are predicted to possess a magnetic moment around 0.3 $\mu_{\rm B}$  per
wire atom, whereas the bulk leads remain nonmagnetic.
In contrast, thicker coaxial (6,1) wires were found to be nonmagnetic.

Of course, thermal fluctuations, very large in a nanosystem, will generally 
act to destroy static magnetic order in the absence of an 
external field. 
There are nonetheless two different fluctuation regimes: slow and fast.  
Slow fluctuations transform a nanomagnet to a superparamagnetic 
state, where magnetization fluctuates on a long time scale
between equivalent magnetic 
valleys, separated, e.g., by anisotropy-induced energy 
barriers. If the barriers are sufficiently large,
the nanosystem spends most 
of the time in a single magnetic valley, and will for many 
practical purposes behave as magnetic. We may under these circumstances
be allowed to neglect fluctuations altogether, and to 
approximate (as we will do here) the calculated properties of the 
superparamagnetic nanosystem with those of a statically magnetized one.
Experimentally, evidence of one-dimensonal (1D) superparamagnetism  with fluctuations
sufficiently slow on the time scale of the probe has been reported
in Co atomic chains deposited at Cu surface steps~\cite{gambardella2002}.
At the opposite extreme --- a situation reached for example 
at high temperatures --- the energy barriers are so readily
overcome that the magnetic state is totally washed away
by fast fluctuations, leading to a conventional paramagnetic state, 
in which all magnetic exchange 
splittings of the electronic bands vanish through 
motional narrowing. 
A complete description of this high entropy state is 
beyond scope, but a conventional $T=0$ nonmagnetic,
singlet solution of the Kohn-Sham electronic structure equations can  
be used in its place, at least as a crude approximation.

The present density functional calculations~\cite{dft} were performed with an all-electron full-potential linear
muffin-tin orbital (FP-LMTO) basis set~\cite{wills} together with 
a generalized gradient approximation (GGA)~\cite{gga} to the 
exchange-correlation functional. 
A simpler local density functional~\cite{lda} was also tested, giving results very similar
the GGA.
As a double check, some of the 
calculations were repeated using the linear augmented plane-wave 
code WIEN97~\cite{wien97}. 
None of these calculations 
assume any shape approximation of the potential or wave functions.
We performed both scalar relativistic calculations, and calculations including the spin-orbit
coupling as well as the scalar-relativistic terms. 
In the calculations involving the spin-orbit interaction, the spin axis was chosen to be 
aligned along the wire direction.

The calculations were performed with inherently three-dimensional
codes. Thus, the infinitely long, straight, isolated monowires, as well as the (6,1) coaxial wires,
were simulated by regular arrays of well spaced nanowires. 
The short wire consisted of three atoms in a straight line attaching to thick planar slabs of close-packed bulk Pd. 
The bond length in the short wire was chosen to be 2.7 {\AA}, as suggested by transmission electron microscope 
images of Pd monowires~\cite{rodrigues}, and 
corresponds to a somewhat stretched wire.
Convergence of the magnetic moment was checked with respect to k-point mesh density, Fourier mesh density,
tail energies, wire-wire vacuum distance and bulk thickness.
In the calculations, spin-orbit coupling was seen to have only a very minor effect on the results, in 
contrast to the situation for $5d$ metals\cite{delin_wires}.

\begin{figure}[h]
\psfig{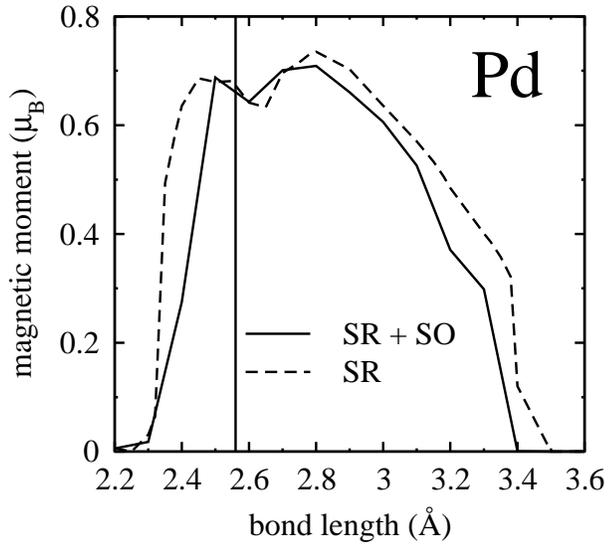}
 \caption
  {
  Magnetic moment per atom as a function of bond length for a long, monatomic Pd wire.
  The vertical line points out the equilibrium bond length. 
  (SR + SO = calculation including spin-orbit coupling and scalar relativistic terms; SR = calculation 
  including scalar relativistic terms but not spin-orbit coupling )
 \label{fig:magnetic_moment}
  }
 \end{figure}

Fig.~\ref{fig:magnetic_moment} shows the magnetic moment per 
atom for the (infinitely) long Pd monowire,
as a function of the bond length along the wire. The total magnetic 
moment rapidly reaches 0.7\,$\mu_B$ at a bond length around 2.3\,{\AA}, and
retains this magnetic moment over a long region.
The steep rise at the onset of magnetism is almost entirely due to $4d$
polarization, which reaches a maximum of $\sim$0.5\,$\mu_B$ 
already at 2.6\,{\AA} and then decreases monotonically.
The remainder of the magnetic moment has $5s$ character.
At around 3.5\,{\AA}, the magnetic moment disappears completely, 
where the wire undergoes a metal-insulator transition, with
the opening of a $ds$ gap, foreshadowing that of the noninteracting
atoms.

For the long monatomic wire, the energy gain per atom due to spin 
polarization is around 25\,meV at the equilibrium bond length 2.56\,{\AA}.
For comparison, we mention that in bulk Ni the corresponding energy 
gain is around 40 meV, indicating that the Pd wire magnetism might actually exist 
not only at ultra-low temperatures.
Antiferromagnetic Pd monowire configurations were also 
tested, and found to be energetically unstable
compared to the ferromagnetic configuration. 
Our result differs from that of Bahn {\it et al.}\cite{bahn2001} who
found no magnetism in pseudopotential calculations for Pd monowires.
It is possible that the disagreement could arise in this very borderline case due to the different methods
used, in which case we would tend to trust our all-electron approach better.

Why does the Pd monowire magnetize?
In the bulk, and also at surfaces,
the Pd $4d$ band is too wide to provoke spin-polarization. In the atom, on the other hand, 
Pd forms a singlet with a completely filled $4d$ shell.
Thus, the limiting cases are all nonmagnetic and the strong Hund's rule magnetism in the wire appears rather unexpected.
However, the borderline case of Pd magnetism is demonstrated by the fact that certain clusters
are predicted to have a magnetic ground state~\cite{landman2001}.
From the atomic perspective, as the distance between Pd atoms decreases,
the increased interatomic hybridization causes the $4d$ band to become partially 
unfilled due to $d \rightarrow s$ transfer, and thus symmetry breaking through
spin-polarization becomes possible. Hund's rules make it reasonable to assume 
that spin polarization will also be energetically favorable. 
From the bulk perspective, the reduction of the number of nearest 
neighbors in the wire compared to the bulk
causes a narrowing of the $4d$ band, and the band width may become 
sufficiently small that the gain in exchange energy due to spin polarization is larger than the increase in kinetic energy.
Thus, using this $4d$ band-narrowing argument, we can rationalize the existence of a magnetic state in the wire.

However, we have reason to believe that
this explanation of the large magnetic moment in long Pd monowires is too simple, and that
$4d$ band-narrowing is not the only mechanism at work, but that the one-dimensionality of the system
is crucial. We will illustrate this point by analyzing the band structure of the long Pd monowires in some detail. 

\begin{figure}[h]
\psfig{figure=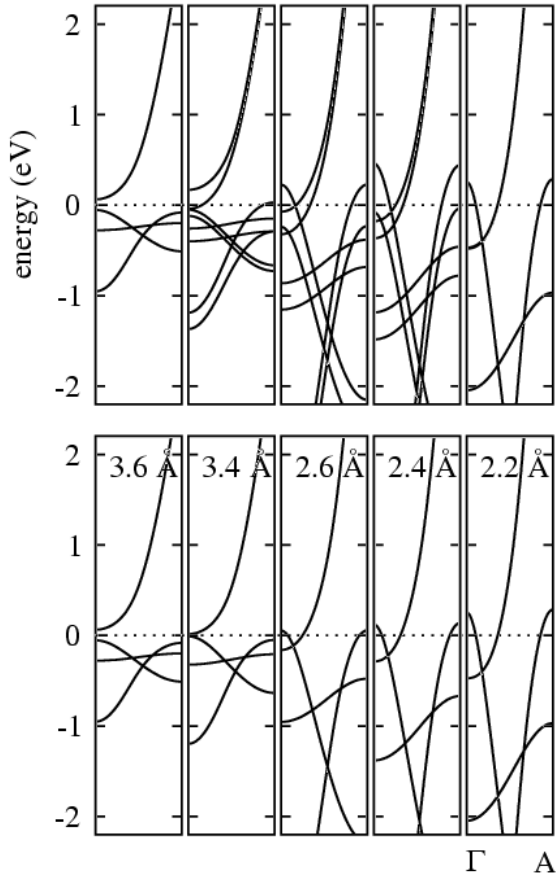,width=8.0cm}
 \caption
  {
  Scalar-relativistic Pd monowire band structures, along the wire direction.
  The Fermi energy is at zero. 
  The upper panels shows the ferromagnetic calculation, and the lower panels the 
  nonmagnetic calculation.
 \label{fig:band_structure}
  }
 \end{figure}

\begin{figure}[h]
\psfig{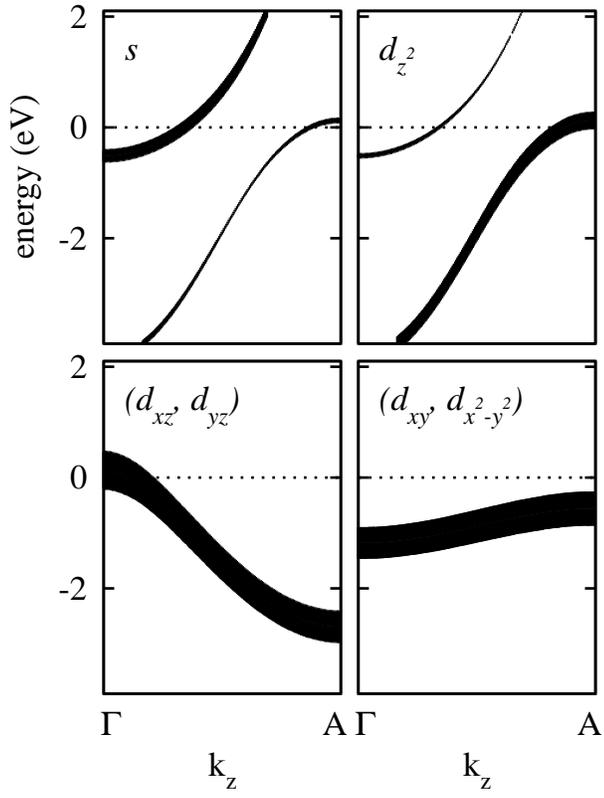}
 \caption
  {
  Character-resolved scalar-relativistic band structure along the wire direction, for a nonmagnetic 
  Pd wire with a bond length of 2.6 {\AA}.
  The Fermi energy is at zero.
 \label{fig:fatbands}
  }
 \end{figure}

Fig.~\ref{fig:band_structure} shows the band structure of the Pd monowire for several different bond lengths,
as it goes from insulator, to ferromagnet, to paramagnet, for decreasing bond length.
There are six orbitals to consider,  
two $s + d_{z^2}$ orbitals, two  $(d_{xz},d_{yz})$ orbitals, and two $(d_{xy},d_{x^2 - y^2})$ orbitals,
as illustrated in Fig.~\ref{fig:fatbands}.
The two $s + d_{z^2}$ orbitals give rise to the two bands with 
highest dispersion. Of these, the band with mostly $s$ character 
has an electron band edge close to the Fermi level at $\Gamma$, 
and the $d$-character dominated one has a hole band edge close to 
the Fermi level at the zone boundary A.
Next, we have the two $(d_{xz},d_{yz})$ orbitals, which are degenerate 
and have a band edge close to the Fermi level at $\Gamma$.
Finally, the two $(d_{xy},d_{x^2 - y^2})$ orbitals form a flat 
band situated between 0.5 and 1\,eV below the Fermi level.
 
The $(d_{xy},d_{x^2 - y^2})$ band lies far below the Fermi 
level with both spins completely occupied, and gives no contribution 
to magnetism. The spin moment is related to the  $s + d_{z^2}$ and  
$(d_{xz},d_{yz})$ bands, which have a high dispersion, and
display one-dimensional band edges close to the Fermi level at $\Gamma$ and A.
In the magnetic regime, i.e., from 2.3 {\AA} to 3.5 {\AA}, these three band edges all are nearly degenerate 
in energy and close to the Fermi level.  
This accidental feature 
of the long Pd monowire band structure dramatically increases the density 
of states, with divergent van Hove singularities close to 
the Fermi level. A spin splitting of these band edges
can thus lower the total electron band energy, somewhat analogous
to a band Jahn Teller, or a Peierls instability effect.  

The fact that the Stoner stability criterion against 
magnetism is necessarily violated leading to a magnetized
nanowire when a band edge approaches
to the Fermi level is universal, as pointed out earlier
by Zabala {\it et al.}~\cite{zabala1998}. However, whereas for a simple $sp$ metal the
ensuing 1D magnetism will imply very small moments
which will readily be washed away
by fluctuations, here the situation is different. The moment is
large, and we expect a robust superparamagnetic state for 
temperatures not too high, transformable to a genuine 
magnetic state under an external field. The 1D singularities ---
in scattering terms the relevant band edge singularities 
correspond to standing waves, here of $s + d_{z^2}$ and  
$(d_{xz}, d_{yz})$ electrons and holes --- provide here merely the mechanism that 
triggers awake a strong Hund's rule magnetic moment. 

If the Pd monowire magnetism is to the most part due to a 1D scattering phenomenon, it
should disappear or at least become much smaller when the number of nearest neighbors becomes large.
To test this assertion, we, as already mentioned, also performed calculations for a two-layer coaxial (6,1) wire.
As expected, it was found to be nonmagnetic,
supporting the view that the magnetism in the Pd monowire is, to a large extent, a 1D phenomenon.
This 1D magnetism in Pd proves nevertheless to be remarkably robust, as it survives even down to three-atom
chains suspended between nonmagnetic leads.

The magnetic moment in the short wire suspended between bulk Pd leads is substantially smaller than the 
moment of the long monowire moment; 0.3 $\mu_B$ per wire atom compared to 0.7~$\mu_B$. 
Due to the high complexity of the band structure for the short wire (not shown), 
it is not possible to single out 
band edges as responsible for triggering the Hund's rule magnetism in the short-wire case.
However, we find that the bulk leads affect the electronic structure in the short wire 
resulting in a wider $4d$ band than in the long
wire, which in turn reduces the magnitude of the magnetic moment.

In summary, we predict that ultimately thin Pd nanowires, 
both short and long, exhibit a spin-polarized ground state.
The resulting superparamagnetic state of the nanowire should
show up in the ballistic conductance in the form of a strong
and unusual magnetic field and temperature dependence as well as
spin-polarization of the current through the wire.
Rodrigues {\it et al.}\cite{rodrigues} recently measured the charge conductance of Pd nanocontacts and
found features in the conductance histogram above as well as below one conductance quantum $G_0 = 2e^2/h$. 
The existence of features below $G_0$ is intriguing since it seems to suggest that the $s$-dominated band crossing the 
Fermi level is spin-split around the Fermi level. A speculative explanation could entail
some kind of spin reversal amid the nanowire~\cite{smogunov2002}.
More theory work will be needed to address their data,
explicitly including such elements as tips, spin structures, and temperature as well
as their effects on the system's conductance. We are currently 
working to address this issue.

\acknowledgments
A.D. acknowledges financial support from
the European Commission through contract no. HPMF-CT-2000-00827 Marie Curie fellowship,
STINT (The Swedish Foundation for International Cooperation in Research and Higher Education),
and NFR (Naturvetenskapliga forskingsr{\aa}det).
Work at SISSA was also sponsored through TMR FULPROP, MUIR (COFIN and FIRB RBAU01LX5H) and by INFM/F.
J. M. Wills is acknowledged for letting us use his FP-LMTO code.
We are also grateful to D. Ugarte for sharing with us the unpublished results of Ref.~\onlinecite{rodrigues}.


%

\end{document}